# Modulation Mechanism of Ionic Transport through Short Nanopores by Charged Exterior Surfaces †

Long Ma,[a,b] Zhe Liu,[a] Jia Man,[a] Jianyong Li,[a] Zuzanna S. Siwy[c] and Yinghua Qiu[a,b,d]*

a. Key Laboratory of High Efficiency and Clean Mechanical Manufacture of Ministry of Education, National Demonstration Center for Experimental Mechanical Engineering Education, School of Mechanical Engineering, Shandong University, Jinan, 250061, China

b. Shenzhen Research Institute of Shandong University, Shenzhen, 518000, China

c. Department of Physics and Astronomy, University of California, Irvine, California 92697, United States

d. Suzhou Research Institute of Shandong University, Suzhou, 215123, China

†. Electronic supplementary information (ESI) available. See electronic supplementary information for simulation details, additional simulation results of ionic current through nanopores, and ion concentration distributions inside nanopores.

*Corresponding author: yinghua.qiu@sdu.edu.cn

**Abstract**

Short nanopores have various applications in biosensing, desalination, and energy conversion. Here, the modulation of charged exterior surfaces on ionic transport is investigated through simulations with sub-200 nm long nanopores under applied voltages. Detailed analysis of ionic current, electric field strength, and fluid flow inside and outside nanopores reveals that charged exterior surfaces can increase ionic conductance by increasing both the concentration and migration speed of charge carriers. The electric double layers near charged exterior surfaces provide an ion pool and an additional passageway for counterions, which lead to enhanced exterior surface conductance and ionic concentrations at pore entrances and inside the nanopore. We also report that charges on the membrane surfaces increase electric field strengths inside nanopores. The effective width of a ring with surface charges placed at pore entrances ($L_{cs}$) is considered as well by studying the dependence of the current on $L_{cs}$. We find a linear relationship between the effective $L_{cs}$ and the surface charge density and voltage, and an inverse relationship between the geometrical pore length and salt concentration. Our results elucidate the modulation mechanism of charged exterior surfaces on ionic transport through short nanopores, which is important for the design and fabrication of porous membranes.

**Introduction**

Recent nanofabrication techniques allow the creation of various nanopores, spanning from polymer nanopores[1] with tens of micrometers in length to two-dimensional nanopores with an atomic thickness.[2] When nanopores are placed between two reservoirs to form an ionic current loop, these nanofluidic devices offer a versatile platform for investigating the transport of ions and fluids in confined spaces, which can shed light on various applications of nanopores, such as biosensing,[3, 4] ionic circuits,[5] desalination,[6, 7] and energy conversion.[8-10]

Under the highly confined space of nanopores, interfacial effects, in particular surface charges at the solid-liquid interface, provide significant modulation of ion transport,[11] which originates from the electrostatic interactions among surface charges and mobile ions.[12] In long nanopores, charged inner-pore surfaces usually provide the most significant influence on the transport of ions and charged biomolecules,[11] through the formation of electric double layers (EDLs). EDLs alter the ionic concentration distributions inside nanopores and enable the ionic selectivity of nanopores.[13, 14]

Due to the strong selective transport of counterions in long nanopores, the ionic concentration becomes enriched and depleted at the exit and entrance of the nanopore, respectively. This phenomenon, known as the ion concentration polarization (ICP),[15] has many potential applications in the preconcentration of charged analytes,[16] and material separation.[17] ICP leads to S-shaped current-voltage (I-V) curves, indicating that a limiting current is reached under relatively high voltages.[18] In these cases, ICP affects the ionic current through the modulation of the ionic concentration at one entrance and inside nanopores, which can become more significant in short nanopores because of the enhanced electric field strength.[19]

Because surface charges of inner-pore walls, instead of those on the membrane surface, determine the properties of ion current through the pore with a high ratio of the pore length to the diameter, in earlier simulations[20-22] and theoretical analysis,[13, 23, 24] the exterior surface charges of nanopores have seldom been considered. As pore length decreases to the nanoscale, due to the smaller length-diameter ratio, the charged exterior surface of the nanopore provides a stronger influence on ionic transport than that of the charged inner surface.[25-27] With the modification using polymer molecules on exterior membrane surfaces to reach different charge polarities, Tagliazucchi et al.[25] realized an ultra-thin current rectifier using a 15 nm in diameter nanopore with 20 nm in length. In nanofluidic simulations with nanopores of various lengths, Ma et al.[26] found that exterior charged surfaces dominated the total ionic current rectification when the length-diameter ratio was less than ~3.5. Using nanopores of 20 nm in length and 8 nm in diameter functionalized with bipolar polyelectrolyte brushes on the inner and exterior surfaces, Lin et al.[27] found that bipolar functional groups carried on exterior surfaces exhibited stronger ability to induce ion current rectification than charges on inner surfaces.

Charged exterior pore walls modulate ionic transport usually through the formation of ion pools at the pore entrance and additional passageways at the pore exit for ionic transport.[19, 28] It was also found that ions in EDL might even exhibit enhanced ionic mobility compared to the ions at the pore center.[29] In our earlier work with a conical nanopore of 100 nm in length, 2 and 20 nm in radii of the tip and base openings, surface charges on the exterior surface were shown to attract a substantial amount of additional counterions, which provided an ion pool and resulted in current rectification.[19] From the simulation of osmotic energy conversion under natural salt gradients,[30] the charged exterior surface on the low-concentration side could significantly promote the diffusion of counterions, which was mainly attributed to the tangential ionic transport along the exterior charged surface.[28, 31]

With the development in nanofabrication and surface modification techniques, short nanopores of controlled dimensions and surface properties can find a variety of different applications, because they offer higher resolution in detection,[32] and larger throughput.[28] In this study, modulation of charged exterior surfaces of short nanopores on ionic transport properties was systematically investigated through simulations with the adjusted surface charges. We show that for nanopores with sub-200 nm length, charged exterior surfaces significantly influence ionic current, with improved cation and anion currents. Based on the detailed analysis of the ionic flux, electric field strength, and fluid flow, the modulation mechanism of charged exterior surfaces on ionic current was determined. Through adjusting the charged area on the membrane surface, the effective width ($L_{cs}$) of the charged ring region near the nanopore was found which may be useful in the propagation of ionic transport from single nanopores to porous membranes. The effective magnitude of $L_{cs}$ was explored under different parameters of the nanopore and applied conditions. With the quantitative relationship between $L_{cs}$ and the pore length, diameter, surface charge density, applied voltage as well as salt concentration, our results provide important guidelines for the design of nanopore arrays in thin porous membranes.

**Simulation Details**

Nanofluidic simulations were conducted with COMSOL Multiphysics. As shown in Figure S1, nanopores connect two reservoirs with 5 μm in radius and 5 μm in length. The diameter of nanopores varied from 4 to 55 nm, and the nanopore length was changed from 20 to 1000 nm to investigate the ionic transport through nanopores under different length scales.[28] Default pore diameter and length were 10 and 50 nm, respectively.[14, 30, 33, 34] The system temperature was maintained at 298 K, and the dielectric constant of water was set to 80. Because KCl solutions are commonly applied in biosensing based on nanopores,[35-37] KCl solutions varying from 10 to 500 mM in concentration were chosen in the simulations, of which 100 mM was the default concentration. The diffusion coefficients of $K^+$ and $Cl^-$ ions were considered as $1.96 \times 10^{-9}$ and $2.03 \times 10^{-9}$ m$^2$/s, respectively.[38] Because of the symmetric geometry and surface charge distribution of nanopores, positive voltages were applied across nanopores which varied from 0 to 1 V. Ionic current at the default voltage of 1 V was used for data analysis.

In our simulation models, we used coupled Poisson-Nernst-Planck and Navier-Stokes equations (Equations 1-4),[19, 32] which describe the ionic distributions near charged surfaces, ionic transport in aqueous solutions, and fluid flow under confined spaces.

$$\varepsilon \nabla^2 \varphi = -\sum_{i=1}^{N} z_i F C_i \quad (1)$$

$$\nabla \cdot \mathbf{J}_i = \nabla \cdot \left( C_i \mathbf{u} - D_i \nabla C_i - \frac{F z_i C_i D_i}{RT} \nabla \varphi \right) = 0 \quad (2)$$

$$\mu \nabla^2 \mathbf{u} - \nabla p - \sum_{i=1}^{N} (z_i F C_i) \nabla \varphi = 0 \quad (3)$$

$$\nabla \cdot \mathbf{u} = 0 \quad (4)$$

in which $\varphi$ and $N$ are the electrical potential and number of ion types. $F$, $R$, $T$, $\mu$, and $p$ are the Faraday's constant, gas constant, temperature, liquid viscosity, and pressure. $\varepsilon$ is the dielectric constant, and $\mathbf{u}$ is the fluid velocity. $\mathbf{J}_i$, $C_i$, $D_i$, and $z_i$ are the ionic flux, concentration, diffusion coefficient, and valence of ionic species $i$ ($K^+$ and $Cl^-$ ions), respectively. The ionic current ($I$) under different voltages was obtained by integrating the ionic flux of cations and anions over the boundary of the reservoir with Equation 5.

$$I = \int_S F \left( \sum_i^2 z_i \mathbf{J}_i \right) \cdot \mathbf{n} \, dS \quad (5)$$

where $S$ represents the reservoir boundary, and $\mathbf{n}$ is the unit normal vector.

To understand the influence of charged exterior surfaces on the ionic properties of nanopores, simulation models of nanopores with uniformly charged surfaces (ACS) and only charged inner-pore surfaces (ICS) have been built. The surface charge density varied from −0.04 to −0.16 C/m$^2$, of which −0.08 C/m$^2$ was used as the default value.[32, 39, 40] The same mesh strategy as used in our previous publications was applied (Figure S2).[19, 28, 41] For charged surfaces, the mesh size of 0.1 nm was applied which is sufficient to fully consider the influence of EDLs on the transport of ions and liquids. The convergence of ionic current through the nanopore was used to determine the minimum size of mesh on pore walls (Figure S2). More simulation details are provided in Table S1.

**Results and Discussion**

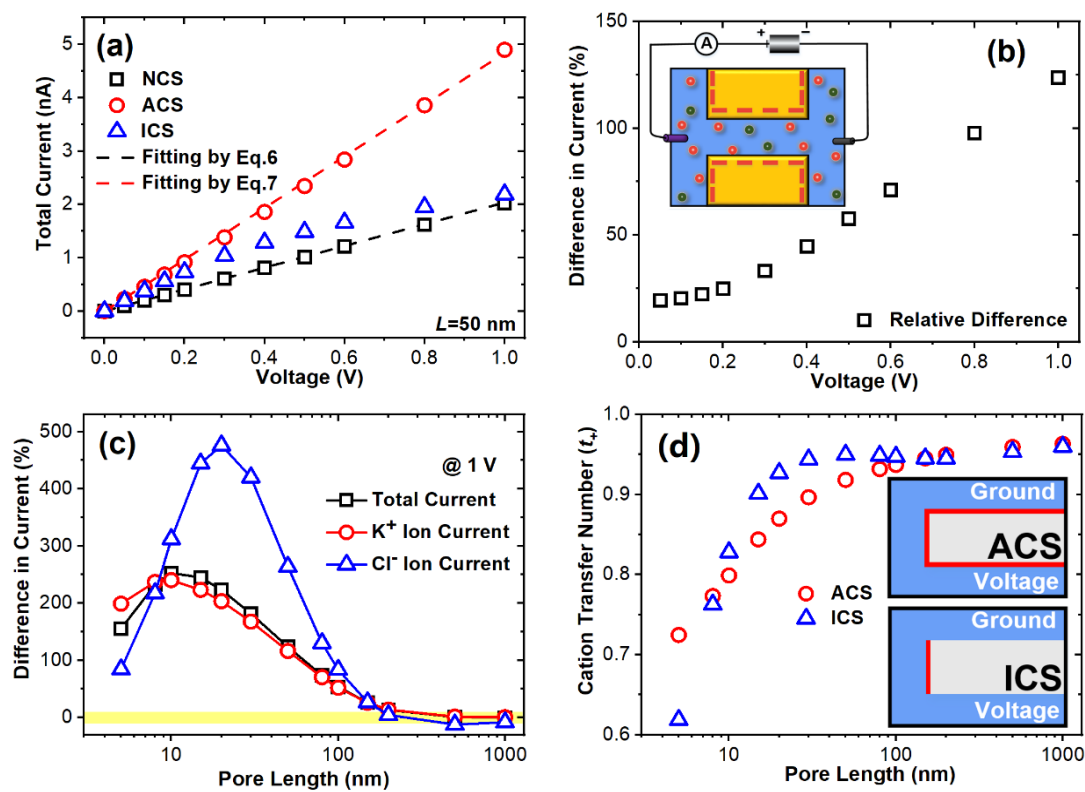

Figure 1. Ionic behaviors in nanopores with all charged surfaces (ACS) and charged inner-pore surfaces (ICS) under different voltages and pore lengths. (a) Current-voltage curves. Nanopores with all neutral surfaces are denoted as the NCS case. Black and red dash lines are the theoretical fitting with Equation 6 and Equation 7, respectively. (b) Difference in ionic current under various voltages calculated with $(I_{ACS} - I_{ICS})/I_{ICS}$. The inset shows the simulation scheme. (c) Difference in ionic current contributed by total ions, $K^+$ ions, and $Cl^-$ ions under various pore lengths calculated as $(I_{ACS} - I_{ICS})/I_{ICS}$. The yellow region indicates the difference in current within ±10%.

(d) Cation transfer numbers in ACS and ICS cases under different pore lengths. Insets show the simulation models of ACS and ICS. Surface charges are shown in red. The pore diameter is 10 nm.

Short nanopores with thicknesses ranging from 10 to 100 nm[42-44] are usually employed for the detection of biological molecules,[3] such as DNA, RNA, and protein molecules, bio-nanoparticles,[45, 46] like liposomes, and viruses, as well as artificial particles of varying sizes.[47, 48] Here, 10-nm-in-diameter nanopores with 50 nm in length were first chosen to investigate the influence of charged exterior nanopore surfaces on ionic transport inside the pores.

Figure 1a shows the I-V curves obtained in nanopores with different charged surfaces. Three simulation cases were considered, i.e. nanopores with all neutral surfaces, NCS, nanopores with only charged inner surfaces, ICS, and nanopores with all charged surfaces, ACS. From the classical theory, nanopores can be considered a resistor in the ionic circuit. For the NCS case, the absence of surface charges results in a uniform ion concentration inside the nanopore, which does not change with the applied voltage. The I-V curve in the NCS case is linear which can be well-fitted using Equation 6 with the consideration of the pore resistance and access resistance.[35]

$$I = V/R_t = V/(R_p + 2R_{ac}) = V/(\frac{4L}{\pi \kappa d^2} + \frac{1}{\kappa d}) \qquad (6)$$

where $V$ is the applied voltage. $R_t$, $R_p$, and $R_{ac}$ are total resistance, pore resistance, and access resistance, respectively. $\kappa$ is the conductivity of the aqueous solution. $L$ and $d$ and the length and diameter of the nanopore.

In the ICS case, the inner-pore surface is negatively charged at $-0.08$ C/m$^2$, such that an EDL layer is formed near the charged surface. Due to the accumulation of counterions in the EDLs, the ion current in the ICS case is enhanced compared to the current through a neutral pore. The ICS and NCS nanopores also differ in their I-V characteristics. At the low voltage regime, below 0.2 V, an ICS nanopore exhibits a linear I-V curve. However, as the voltage increases, the I-V curve becomes S-shaped. The nonlinearity of the pore behavior is caused by concentration polarization (ICP) that induces the formation of a depletion zone at one pore entrance where the concentration of both cations and anions is lower than in the bulk.[15] ICP results from ion selectivity and becomes more pronounced at higher voltages. As the length of a nanopore increases, ICP gradually weakens at the same applied voltages. When the nanopore length exceeds 200 nm, I-V curves in the ICS case show a linear characteristic in the whole examined voltage range between 0 to 1 V (Figure S3).

As mentioned above, in many previous simulations, the exterior surface charges of nanopores were ignored, particularly when nanopores had micrometer-scale lengths.[20-22] Recent studies[25, 26] show that as the pore length shortens, the exterior surface charge may have a significant impact on the ion current inside the nanopores. Here, charged exterior surfaces are considered in the ACS case. The I-V curve for the ACS case is almost linear, which can be fitted using Equation 7 referring to the work of Lee et al.[49] with the consideration of the surface charge effect on the pore resistances and access resistances.

$$I = V \cdot G = V \cdot \kappa \cdot \left[ \frac{4L}{\pi d^2} \frac{1}{1 + 4\frac{l_{Du}}{d}} + \frac{2}{\alpha d + \beta l_{Du}} \right]^{-1} \quad (7)$$

where $G$ is the nanopore conductance. $\alpha = 2$ is a geometrical prefactor, and $\beta = 2$ is a numerical constant. $l_{Du}$ is the Dukhin length, $l_{Du} = \kappa_s/\kappa_b$ with $\kappa_s$ and $\kappa_b$ the surface conductivity and bulk conductivity, respectively.

The good fit between Equation 7 and the simulation result also indicates that the surface charges in the ACS case affect both the ionic transport inside the nanopore and at the nanopore opening. Note that although ICP is still present in the ACS case (Figure 2a), the limiting current does not appear in the I-V curve, which will be explained later. The current values in the ACS case are ~2.5 times higher than those in the NCS case. Compared with the ICS system, ionic transport in the ACS nanopore is greatly improved as well. Figure 1b shows the relative current difference in both ACS and ICS cases, which becomes more pronounced with applied voltages. At 1 V, the presence of charges on the membrane exterior increases ion current by ~125%.

In Figure 1c, the current difference between ACS and ICS systems at 1 V is analyzed as a function of pore length. The current difference is calculated as ($I_{ACS}$-$I_{ICS}$) / $I_{ICS}$, where $I_{ACS}$ and $I_{ICS}$ represent the ion current obtained from the ACS and ICS cases, respectively. The pore diameter is kept at 10 nm, and the pore length varies from 5 to 1000 nm. When the pore length exceeds 200 nm, the difference in currents is less than 12% and the influence of exterior surface charges can be ignored. However, as the pore length decreases to sub-200 nm, the current difference between ICS and ACS cases, including cation current, anion current, and total current, increases significantly with further shortening of the pore length. In nanopores with a length of ~10-20 nm, the current difference between both cases reaches a maximum that can exceed ~240%. Here, we focus on the total current and $K^+$ ion current because of the cationic selectivity of the negatively charged nanopores. When the pore length is further reduced to 5 nm, the difference between total currents decreases but remains at a high value of ~150%. Based on the comparison between ion currents in both ICS and ACS cases, we conclude the charged exterior surfaces have a significant impact on the transport of cations and anions inside the short nanopores.

As negatively charged nanopores are selective to cations, we also considered a cation transfer number ($t_+$)[14] under various pore lengths. For nanopores with a length of less than 10 nm, $t_+$ in the ACS case is higher than that in the ICS case, Figure 1d. Figure 1d also shows that increasing the pore length from 5 to 100 nm enhances the cation selectivity in both cases, ICS and ACS nanopores, and $t_+$ saturates at ~0.95 at the pore length of ~100 nm. Surprisingly, the cation transfer number in the ACS case is smaller than that in the ICS case, most probably due to an enhanced contribution of Cl$^-$ to ion current, which is discussed below. For nanopores longer than 200 nm, the cation selectivity is determined by the charged inner pore surface, and the influence of the charged exterior surface can be ignored.

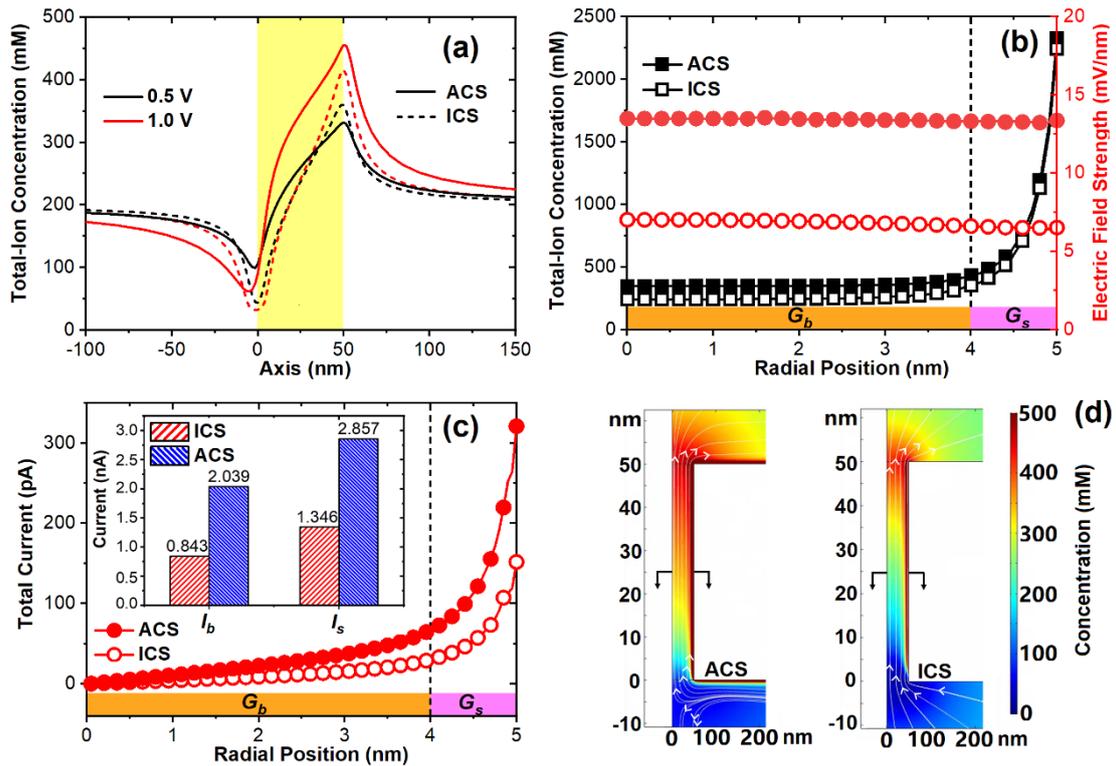

Figure 2. Details of ion transport in both ACS and ICS cases. (a) Concentration distributions of all ions along the pore axis at different voltages. (b) Radial distributions of ion concentration and electric field strength in the center cross-section of the nanopore. The orange and magenta represent the regions of bulk conductance ($G_b$) and surface conductance ($G_s$) inside nanopores, respectively. The surface conductance was determined within 1 nm from the surface. (c) Radial distributions of ionic current at the center cross-section of the nanopore. Each value was obtained through the integration of the ionic flux over a 0.05-nm-wide segment. Inset shows the total bulk current ($I_b$) and surface current ($I_s$) in both cases. (d) 2D concentration

distributions of total ions (color maps) and fluid flow (white lines with arrows). Black arrows show the center cross-section of the nanopore.

To reveal the regulation mechanism of exterior surface charges on ionic transport in short nanopores, we investigated the details of ionic properties of nanopores,[28, 41] such as the distribution of ion concentration, electric field strength, and ion flux, as shown in Figure 2. Figure 2a shows the ion concentration distribution along the pore axis. In both ICS and ACS cases, ICP causes voltage-dependent formation of ionic depletion and enrichment zones at the left and right ends of the nanopore, respectively. In the ACS case, the charged exterior surfaces provide an additional passageway for counterions (Figure 2d), such that more ions can enter the nanopore, resulting in a higher ion concentration than that in the ICS case. Note that in the ACS case, the lowest concentration appears outside of the nanopore. The higher ionic concentration and enhanced surface conductance in the ACS system prevent the formation of the limiting current at higher voltages such that the I-V curves are linear.

Based on the continuity of ion current through the nanopore, the electrochemical properties of the center cross-section of the pore can be used to analyze the impact of the charged exterior surface on the ionic transport quantitatively.[28, 41] Figures 2b and 2c show the radial distributions of the ion concentration, electric field strength, and ionic current at the center cross-section of the nanopore. In the nanopore, the ionic current is directly related to the number and migration rate of charge carriers, which are determined by the ion concentration and electric field strength based on the Nernst-Einstein equation. From the concentration distribution, EDLs form within ~1 nm of the charged inner surface with the accumulation of cations, as predicted from the Debye length.[12] Comparing both radial distributions of ion concentration in the ACS and ICS cases, charged exterior surfaces increase the ion concentration in the center and EDLs regions of the nanopore by ~70% and ~5%, respectively. The electric field strength at the central cross-section shows no obvious radial dependence, which represents almost a constant value at different radial positions. To our surprise, the electric field strength at the central cross-section of the nanopore in the ACS case is roughly twice that in the ICS case. The larger electric field strength can provide the ions with a much higher migration rate. With the charged exterior surfaces of the nanopore, both the ion concentration and ionic migration rate are enhanced, which can result in a significant increase in ionic current.

Figure 2c illustrates the radial distributions of ion current at the central cross-section of the nanopore in both cases of ICS and ACS, which follow a similar trend to that of ion concentration distribution. Due to the higher ion concentration in the EDLs, a larger local ion current appears near the charged inner surface. In charged nanopores, the conductance can be divided into surface conductance in the EDL region near charged surfaces, and bulk conductance in the center region of the nanopore. Considering that the Debye length in 0.1 M KCl solution is ~1 nm, the current contributed by ions within 1 nm from the pore wall and ions 1 nm away from the surface is defined as the surface current $I_s$, and bulk current $I_b$, respectively.[30,49] From Figure 2c, a larger current appears at each position on the cross-section in the ACS case compared to the ICS system, which results in larger bulk and surface currents. As shown in the inset of Figure 2c, the surface current and the bulk current are increased significantly by ~142% and ~112% respectively when exterior surface charges are added.

In aqueous solutions, the directional transport of hydrated ions can lead to the fluid flow i.e. electroosmotic flow (EOF).[14] By including the Navier-Stokes equations in the simulations, the fluid flow characteristics could be captured which can reflect the flow trajectory of ions. As seen in Figure 2, when surface charges are considered on the exterior pore walls, cations accumulate near the charged surface which can be driven by the applied electric field to migrate parallel to the exterior surface. In the ACS case, all the EDLs regions near charged pore walls can serve as an additional passageway of the counterions, which leads to a much larger surface current. The EDLs near the charged exterior surface provide a pool of $K^+$ ions, which can quickly migrate into the nanopore and result in a higher ion concentration at the orifice and inside the nanopore. Note that MD simulations of ionic transport in nanopores revealed a possibility for the counterions to exhibit enhanced mobility in the EDL region compared to the pore center.[29] In the ICS case under high electric field strengths, on the other hand, ion depletion at the entrance of the nanopore becomes more significant due to insufficient ion supply, which leads to the occurrence of limiting current.

It is important to note that besides the enhancement in ionic transport, charged exterior surfaces also promote the velocity of EOF inside the nanopore. Figure S4 shows that the velocity of EOF in the ACS case is ~40% larger than that of the ICS case. The EOF can enhance the transport of counterions through enhancement of the concentration and migration rate of counterions inside the nanopore. Importantly, in the ACS case, more $K^+$ ions accumulate inside the nanopore, leading to better screening of the surface charges by the counterions, allowing more $Cl^-$ ions to be transported through the nanopore. The maximum current of $K^+$ ions and $Cl^-$ ions in the ACS case are increased by up to 112% and 256%, respectively, compared to the ICS nanopore (Figure S5). The enhancement of current due to $Cl^-$ ions is responsible for the decreased transference number of the ACS system shown in Figure 1.

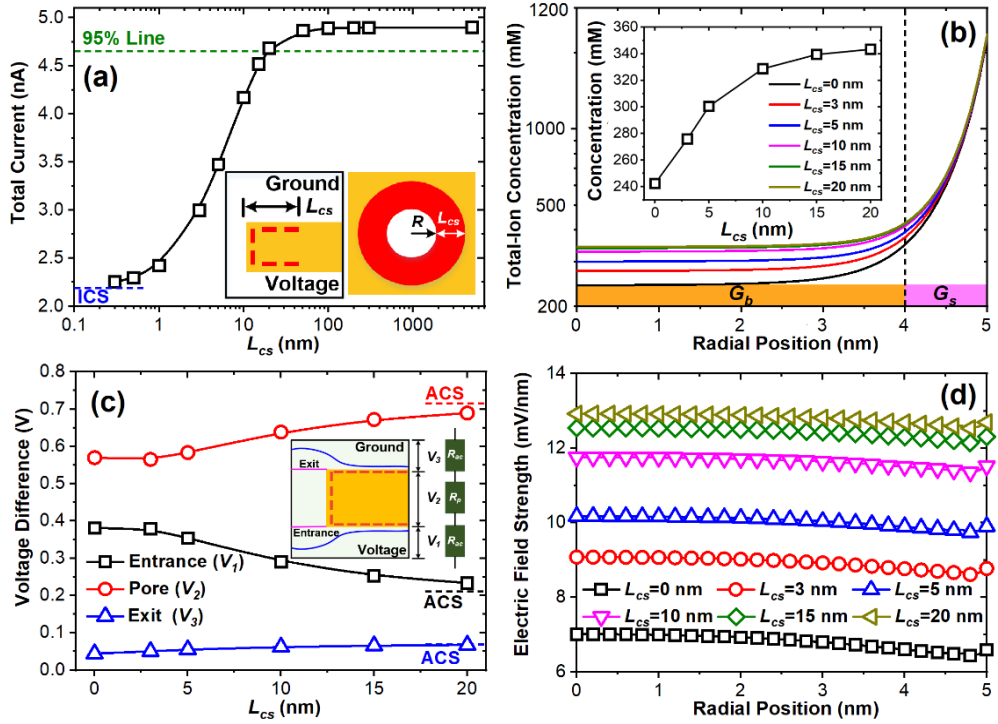

Figure 3. Details of ion transport through a nanopore containing a charged ring at both entrances. The width of the ring is denoted as $L_{cs}$. (a) Enhancement in ionic current by the presence of a charged ring as a function of $L_{cs}$. Inset shows the simulation scheme, in which dashed red lines indicate the surface charges. The diameter and length of the nanopore were 10 and 50 nm, respectively. (b) Radial concentration distributions of all ions in the center cross-section of the nanopore under different $L_{cs}$. Inset shows the ion concentration at the center of the nanopore under different $L_{cs}$. (c) Potential difference across the nanopore and both orifices of the nanopore under different $L_{cs}$. $V_1$ is the voltage between the entrance and the working electrode. $V_2$ is the voltage across the nanopore. $V_3$ is the voltage between the pore exit and the ground. (d) Radial distributions of the electric field strength in the center cross-section of the nanopore under different $L_{cs}$.

In our simulation model, the nanopore is located at the center of the membrane with a diameter of 5 μm. Figures 1 and 2 provide evidence that exterior surface charges significantly influence ionic transport through nanopores. As the next step, we decided to determine the effective length of the charged area outside a pore, considering a ring with surface charges of thickness, $L_{cs}$, placed at both pore entrances which may be important to the design of nanopore arrays. To this end, we conducted simulations of ion current as a function of $L_{cs}$ [28, 30, 31, 41] that varied from 0 to 5000 nm. The two values of $L_{cs}$=0 and $L_{cs}$=5000 nm correspond to the ICS and ACS nanopores, respectively. As shown in Figure 3a, increasing $L_{cs}$ from 0 to ~20 nm leads to an increase in ion current by ~114% from ~2.2 nA to ~4.7 nA. As $L_{cs}$ further increases, the total current gradually saturates at the value of the ACS case. The effective length of the charged ring region on the exterior surface is defined as the value of $L_{cs\_eff}$ at which the total current reaches 95% of that in the ACS case. For the nanopores considered in Figure 3a, the $L_{cs\_eff}$ is ~20 nm under 1 V in 0.1 M KCl.

Ionic current through the nanopore depends on $L_{cs}$, which modulates ion concentration and the electric field strength inside the pore. To understand how the exterior surface charges influence the ionic transport inside nanopores, the radial distributions of ion concentration and electric field strength at the center cross-section of the nanopore were investigated under different values of $L_{cs}$. We found that the ion concentration distributions under different $L_{cs}$ share a similar profile, Figure 3b. The ion concentration in the region beyond the EDLs increases from 242 mM to 343 mM with $L_{cs}$ increasing from 0 to 20 nm. Note that the total resistance of the nanopore system includes the access resistance on both sides of the pore and pore resistance. The applied voltage can then be divided into three parts $V_1$, $V_2$, and $V_3$ (Figure 3c). Ionic distributions in Figure S6a show that an increase in $L_{cs}$ increases the ion concentration at the nanopore entrance, due to the parallel transport of ions to the charged exterior surface, which weakens the ionic depletion phenomenon. The increased ion concentration effectively reduces the access resistance, which in turn induces a larger potential difference across the nanopore ($V_2$). As shown in Figure 3c, with $L_{cs}$ increasing from 0 to 20 nm, the potential differences at the pore entrance ($V_1$) and across the nanopore ($V_2$) decrease and increase by ~40% and ~20%, respectively, which is also observed in the axial distributions of the electric potential (Figure S6b). When $L_{cs}$ reaches 20 nm, the electric field strength at the center cross-section of the nanopore is enhanced by ~86%, from 7 to 13 mV/nm compared to $L_{cs}$ = 0 nm, Figure 3d.

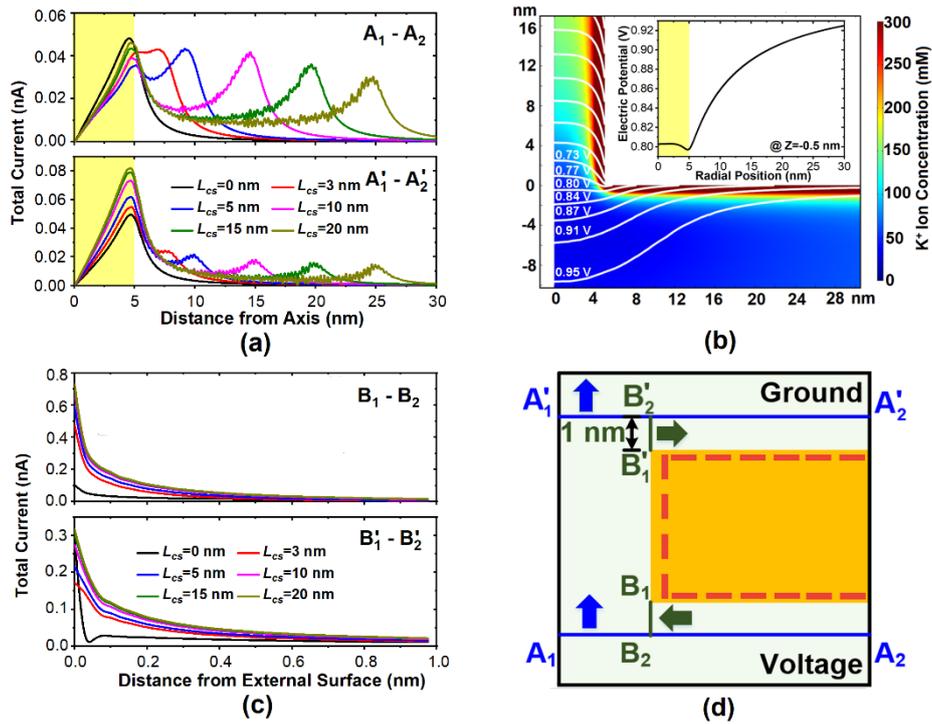

Figure 4. Characteristics of ionic transport through a nanopore under various $L_{cs}$ on both exterior membrane surfaces. (a) Distributions of the total current in the plane located at 1 nm above the membrane surface ($A_1$-$A_2$ and $A'_1$-$A'_2$). (b) 2D concentration distributions of $K^+$ ions (color maps) and equipotential lines (white lines). The inset shows the radial distribution of the electric potential at a distance of 0.5 nm from the exterior surface, i.e. the center of the EDLs. (c) Distributions of the total current along the vertical plane with 1 nm in length on both sides of the membrane ($B_1$-$B_2$ and $B'_1$-$B'_2$). (d) Scheme of locations where current distributions were obtained. $A_1$-$A_2$ and $A'_1$-$A'_2$ represent the plane located at 1 nm above the membrane surface. $B_1$-$B_2$ and $B'_1$-$B'_2$ denote a surface with 1 nm in length along the inner surface located at the pore entrance and exit. Arrows show the directions of the ionic flux. Because the thickness of the EDLs is ~1 nm in 0.1 M KCl, the planes at 1 nm and 0.5 nm above the membrane surface were selected as the boundary and center of EDLs.

We also looked in detail at ionic transport at the pore entrance and exit, where we analyzed the distributions of ion current, see Figure 4. [28, 41] Note that the current distributions were analyzed to show the contribution of external surface conductance to the total current inside the nanopore. $A_1$-$A_2$ and $A_1'$-$A_2'$ represent two planes located at 1 nm above the membrane, around the boundary of the EDLs on both sides of the nanopore. $B_1$-$B_2$ and $B_1'$-$B_2'$ are toroidal surfaces with a length of 1 nm along the inner pore wall, which are perpendicular to the membrane. The distributions of ion current along the $A_1$-$B_2$-$A_2$ and $A_1'$-$B_2'$-$A_2'$ surfaces were obtained. At the pore entrance, counterions can enter the pore directly by passing across $A_1$-$B_2$, or accumulating in the EDL region near the charged exterior surface and then migrating across $B_1$-$B_2$. Similar passageways for ions can be seen at the pore exit.

As shown in Figure 4a, when $L_{cs}$ is 0 nm, i.e. the ICS case, counterions enter the nanopore mainly through the surface $A_1$-$B_2$, and exit the pore at a similar position ($A_1'$-$B_2'$). The current shows a single peak near the charged inner surface of the nanopore, and the ion flux parallel to the exterior surfaces is almost zero. When the exterior surfaces are charged, the EDLs regions provide a fast passageway for ionic transport which induces an obvious surface current. Abundant counterions pass through the $B_2$-$A_2$ plane resulting in a distinct peak of ionic flux near the boundary of the charged surface (Figure 4a). These counterions can enter the nanopore through $B_1$-$B_2$ plane under the voltage drop along the exterior surface, which leads to a considerable ion flux as shown in Figure 4c. In Figure 4b, the radial potential difference at the pore entrance is ~0.15 V which drags ions in the EDLs into the nanopore rapidly. With $L_{cs}$ increasing from 0 to 20 nm the radial potential difference along the charged exterior surface gradually enhances, which can provide an enhanced promotion in the ion flux. On the exit side, most ions leave the pore directly through $A_1'$-$B_2'$ leading to a large current peak. The surface charges on the exterior surface can provide electrostatic attraction to counterions to confine their movement inside the EDLs region. This amount of ions then enter the bulk through the $B_2'$-$A_2'$. Figure 4c shows the ionic current through the $B_1$-$B_2$, and $B_1'$-$B_2'$ planes, which represents the ion flux entering and leaving the nanopore through the EDLs regions. With the increase of $L_{cs}$, both ion fluxes through the $B_1$-$B_2$ and $B_1'$-$B_2'$ increase and reach the maximum at $L_{cs}$ ~20 nm. Since counterions tend to migrate into the bulk directly on the exit side, the ion flux through the $B_1'$-$B_2'$ is lower than that through the $B_1$-$B_2$.

Figure S7 shows the contribution of current through different cross-sections to the total current. When only the inner surface is charged, the ion flux passing through the $A_1$-$B_2$ surface accounts for 60% of the total ion flux. After the exterior surfaces are charged, the rapid transport of counterions along the exterior surface becomes to dominate the total ionic flux. When the $L_{cs}$ reaches 20 nm, the ion current passing through the $B_1$-$B_2$ plane accounts for 75% of the total current. In Figure S7b, surface charges on the exterior surface have a relatively weak influence on ionic transport, which increases the proportion of ion flux along the exterior surface from ~48% to ~53% of the total current.

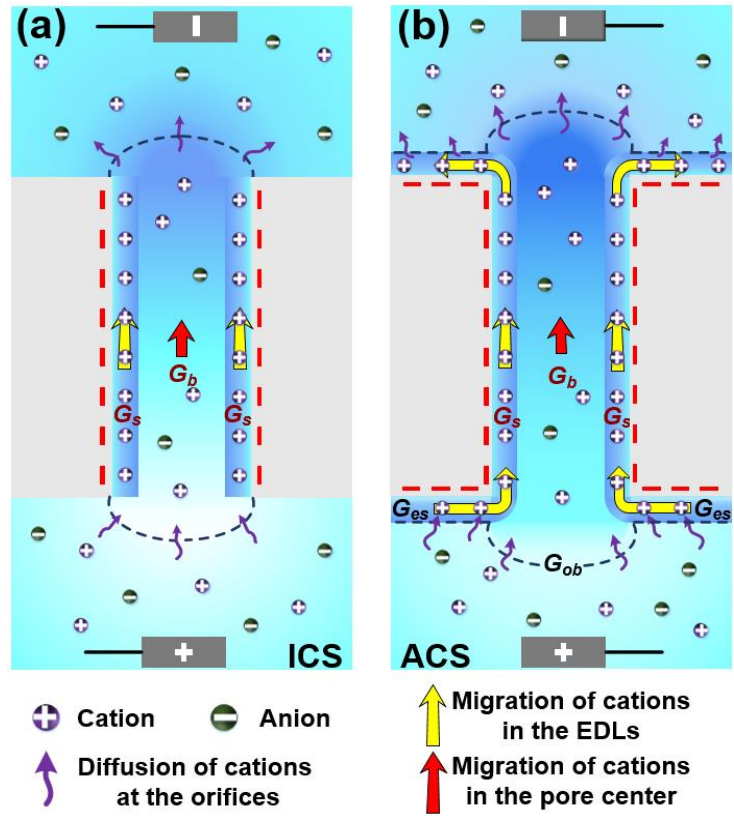

Figure 5. Illustrations of the ionic transport under electric fields in the ICS (a) and ACS (b) cases. Yellow and red thick arrows show the flux of cations in EDLs near charged surfaces and in the pore center, respectively. Curved thin arrows show the ionic flux at the orifices of the nanopore. Dashed red lines indicate the surface charges.

Figure 5 shows the mechanisms of ion transport in ACS and ICS cases under electric fields. The total resistance includes the access resistance at both ends and the pore resistance. Inside the pore, the conductance is composed of the surface conductance near the charged inner wall and the bulk conductance in the pore center. When the exterior surfaces are neutral, counterions mainly transport across the boundary of the access resistance region through ionic diffusion, which then migrate through the pore under electric fields. At the exit end, the counterions diffuse across the boundary of the access resistance region into the bulk. Due to the weak ionic supply by ionic diffusion and the fast ionic migration under applied voltages, ionic depletion and enrichment appear at the pore entrance and exit ends, respectively. Consequently, a depletion zone is created which limits the transmembrane current.

We believe that in the ACS case, the EDLs near charged exterior surfaces of the nanopore not only provide a high-concentration pool of the counterions but also serve as a fast passageway for ionic transport.[19, 30] Thus, in the ACS system, besides the ionic diffusion near the access resistance region, a large number of counterions can enter the nanopore through the migration in the EDLs along the exterior surfaces, which weakens the ionic depletion at the pore entrance effectively. These ions then pass across the pore through the pore center or the EDLs region near the inner surface leading to current increase. On the exit side, the exterior surface charges confine a portion of ionic transport in the EDLs regions through electrostatic interaction. The charged exterior surfaces increase the area for ionic diffusion from the pore to the bulk.

Following the concepts of surface conductance and bulk conductance inside charged nanopores,[30, 49] we propose exterior surface conductance ($G_{es}$) as the ionic flux along charged membrane surface, and orifice bulk conductance ($G_{ob}$) as the ionic migration at the pore entrance. The voltage drop at the pore entrance due to access resistance facilitates sourcing the ions from the bulk, which leads to both exterior surface conductance and orifice bulk conductance.

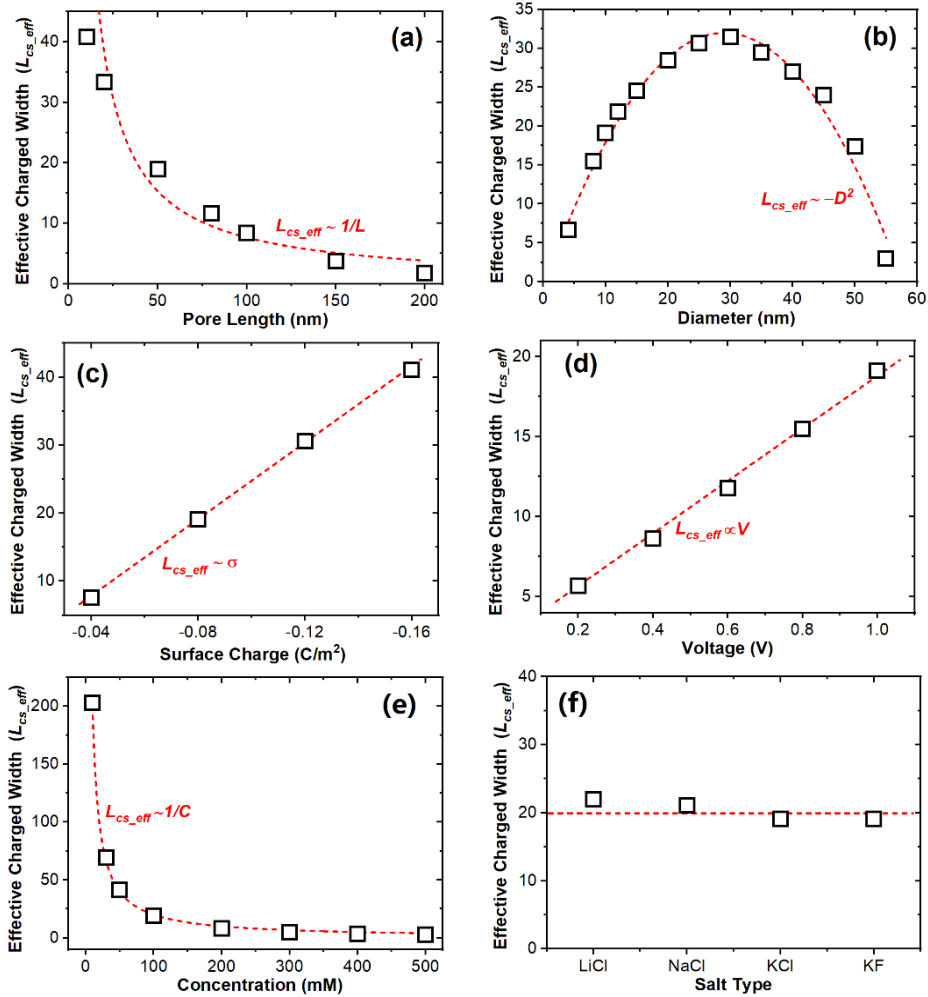

Figure 6. Effective length of the charged ring region on the exterior surface ($L_{cs\_eff}$) under various conditions, such as (a) the pore length, (b) the diameter, (c) the surface charge density, (d) the voltage, (e) the salt concentration, and (f) the salt type.

From Figure 3a, with the charged region around the nanopore expanding, the ionic current through the nanopore can be promoted. When the $L_{cs}$ reaches a certain threshold (~20 nm), i.e. the effective $L_{cs}$, the ion current does not increase with further enlarging of $L_{cs}$. The effective $L_{cs}$ shows the required membrane dimension for independent nanopores. It is important in the propagation of ionic transport through single nanopores to cases with nanopore arrays,[28, 31] which can have many applications in nanofluidic high-throughput detection, high-performance osmotic energy conversion, and seawater desalination. Here, the effective lengths of the charged exterior surface under various conditions have been investigated. Through varying the value of $L_{cs}$, when the ionic current reaches 95% of the current in the ACS case, the corresponding $L_{cs}$ is the effective value ($L_{cs\_eff}$) which represents the required minimum charged surface area.

Figure 6 shows the required $L_{cs\_eff}$ as a function of pore length and diameter, surface charge density, voltage, salt concentration, and salt type. As the nanopore length increases, the magnitude of $L_{cs\_eff}$ decreases following the relationship of $1/L$. The increase in the pore length also decreases the electric field strength inside the pore, which weakens the concentration polarization at both ends of the nanopore. As the pore gets longer, due to the increased pore resistance, the voltage over the pore mouth decreases, which also reduces the ion transport in the EDL regions near the charged exterior surface. For long nanopores, the influence of the charged exterior surface can be ignored (Figure S8a).

For 50-nm-long nanopores with a larger diameter, the access resistance takes a larger portion of the total resistance.[35] The resulting higher potential difference at the pore mouth enhances the migration of counterions in the EDL regions near the charged exterior surface. $L_{cs\_eff}$ increases as the diameter varies from 5 to 30 nm. However, with the diameter further increasing to 50 nm, the orifice bulk conductance gradually dominates the whole conductance, and the promotion of charged exterior surface on ionic current weakens (Figure S8b). Also, because larger pore size decreases the ionic selectivity of the nanopore,[13] the concentration polarization at both ends becomes less significant. The $L_{cs\_eff}$ decreases with the pore diameter.

The surface charge density is another important property of nanopores, which determines the ionic selectivity of nanopores.[13] In nanopores with higher surface charge densities, a higher ion flux due to the larger surface conductance depletes the ion concentration at the pore entrance, such that a larger $L_{cs}$ is required to sustain the ion supply through the exterior surface conductance. Under the surface charge density varying from −0.04 to −0.16 C/m$^2$, the $L_{cs\_eff}$ increases from ~5 to ~40 nm. (Figure 6c).

The influence of applied voltage on the $L_{cs\_eff}$ is similar to that of the surface charge density. Under electric fields, ionic migration provides the main contribution to the total current through a nanopore. As the applied voltage increases, ionic migration enhances inside the nanopore, especially in the EDLs regions near the charged pore wall, which depletes counterions at the pore entrance. As a result, a larger $L_{cs\_eff}$ is required to maintain the large ionic transport. From Figure 6d, the $L_{cs\_eff}$ increases linearly with the applied voltage.

For aqueous solutions used in the work, two properties were considered, i.e. the concentration and salt type. In KCl solutions, the surface charges can be screened more efficiently under a higher salt concentration,[12] which weakens the surface conductance across the nanopore. In this case, the bulk conductance dominates the total ionic flux, resulting in a shorter $L_{cs\_eff}$. From Figure 6e, $L_{cs\_eff}$ is inversely proportional to the solution concentration. The thickness of the EDLs i.e. the Debye length λ is inversely proportional to the square root of the solution concentration (Table S2). On the logarithmic scale, the $L_{cs\_eff}$ and λ satisfy a linear relationship (Figure S9). Finally, we looked into the dependence of $L_{cs\_eff}$ on the type of salt, which necessitated considering different values of diffusion coefficients. Here, LiCl, NaCl, KCl, and KF are used to probe the effect of different counterions and coions on $L_{cs\_eff}$. The ion diffusion coefficients follow the order of Cl$^-$>K$^+$> F$^-$>Na$^+$>Li$^+$.[38] As shown in Figure S8f, the differences in currents obtained in ACS and ICS pores in the four salt solutions follow a similar trend, i.e. they first increase and then remain constant as $L_{cs}$ increases. Figure 6f shows that the $L_{cs\_eff}$ remains almost constant in salt solutions with different diffusion coefficients.

## 4. Conclusions

Systematical simulations have been conducted to investigate the influence of charged exterior surfaces on ionic transport through nanopores. We show that the EDLs near the charged membrane surfaces provide an ion pool and an additional passageway for counterions, which can migrate along the EDLs into the nanopore under the potential difference parallel to the exterior surface. This exterior surface conductance weakens the ion dissipation at the pore entrance but enhances the ionic enrichment inside the nanopore. Compared with the ICS case, charged exterior surfaces improve the ion concentration in the central and EDLs regions at the center cross-section of the nanopore by ~70% and ~5%, respectively. The electric field strength at the center cross-section is also increased by ~100%. The combination of the increase in ion concentration and the enhancement in ionic migration rate results in a 125% increase in the ion current in the ACS case than that in the ICS case. With the adjustment of the charged membrane area, the effective $L_{cs}$ is found ~20 nm. In addition, quantitative relationships between the effective $L_{cs}$ and the nanopore parameters and applied conditions are explored. The effective $L_{cs}$ has a reciprocal relationship with the pore length and salt concentration but is linearly proportional to the surface charge density and voltage. Effective $L_{cs}$ also increases with the pore diameter squared and remains independent of the ionic species. We think the elucidated modulation mechanism of charged exterior surfaces on the ionic transport through short nanopores can provide theoretical guidance for the design and preparation of porous membranes in various applications.

**Author Contributions**

Long Ma: Methodology (equal); Formal analysis (lead); Software (lead); Validation (lead); Writing – Original Draft (lead); Writing – Review & Editing (equal). Zhe Liu: Software (equal); Data Curation (equal); Visualization. Jia Man: Data Curation (equal). Jianyong Li: Investigation (equal). Zuzanna S. Siwy: Supervision (equal); Writing – Review & Editing (equal). Yinghua Qiu:

Conceptualization (lead); Resources (lead); Supervision (lead); Methodology (lead); Investigation (lead); Writing – Original Draft (equal); Writing – Review & Editing (lead); Funding Acquisition (lead).

**Conflict of Interest**

There are no conflicts to declare.

**Acknowledgment**

This research was supported by the National Natural Science Foundation of China (52105579), the Natural Science Foundation of Shandong Province (ZR2020QE188), the Guangdong Basic and Applied Basic Research Foundation (2023A1515012931), the Natural Science Foundation of Jiangsu Province (BK20200234), and the Qilu Talented Young Scholar Program of Shandong University.